# Review Study For Inter-Operability Of Manet Protocols In Wireless Sensor Networks

Gurpreet Singh Saini[#1], Priyanka Dubey[#2], Md Tanzilur Rahman[#3]
[#]Amity School of Engineering and Technology, Amity University
Sec-125, NOIDA, Uttar Pradesh, India

*Abstract*— Wireless Networks are most appealing in terms of deployment over a wide range of applications. The key areas are disaster management, industrial unit automation and battlefield surveillance. The paper presents a study over inter-operability of MANET (Mobile Ad-Hoc Network) protocols i.e DSDV, OLSR, ZRP, AODV over WSN (Wireless Sensor Network) [10]. The review here covers all the prevailing protocol solutions for WSN and deployment of MANET protocols over them. The need of moving to MANET protocols lie in situation when we talk about mobile sensory nodes which are a compulsion when we talk about the above mentioned three areas. However, the deployment may not be limited to these only.

*Keywords*— MANET, WSN, Protocol Inter-operability, QoS Routing, Lifetime based routing, Minimum cost Routing.

## I. INTRODUCTION

There is growing need of mobility in life and hence comes a need of mobile WSN [1]. The area's which may seek this kind of deployment are Battlefield Surveillance, Disaster Management and Industrial unit automation. Sensory nodes are cheap today and have found mass adoption at many stages. The deployment of any protocol over WSN must address following constraints of WSN [1][2]:
- Dense Deployment
- Power Constraint
- Low Computation Power
- Memory Constraint
- High Unreliability due to Ad-Hoc Deployment

When we talk about these scenarios of deployment, we have a situation of criticality where the communication needs to be quick, correct, concise and precise. The infrastructure needs to establish every possible path available according to what they sense. Now, coming onto terms of MANET protocols, they are efficient but have limitations of resolving the constraints levied by the WSN nodes; Power being the prominent one. Studies prove that maximum energy is spent in sensing while less in transmission of the data. So, the MANET protocols need to be modified as per the prevailing WSN protocol development approaches which are:
- Data- Centric
- Energy-Aware
- Sensory and Negotiation based

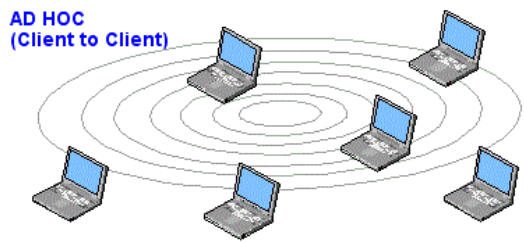

Figure 1: A general Wireless Ad-Hoc Network

The protocols manipulated must also answer following issues:
- Network Dynamics
- Architectural/Design Issues
- Energy Consumption
- Node Deployment Scenario
- Data Transmission Mode (Aggregation / Fusion)
- Node Capabilities / Heterogeneity

Here in this paper we try to resolve issues with view of practical transformation of protocols so as to have a commercially viable output. To do this, we need to take into consideration a scenario of any three usage areas.

## II. BACKGROUND

Let's imagine a scenario of sudden disaster in terms of Earthquake. The earthquake struck suddenly and there was no pre setup to sense the disaster. There are many casualties, infrastructure loss and only thing present is survivors having PDA's with sensing capabilities.

Solution: To this, a solution in terms of a Mobile-Ad Hoc sensory network with following features present in its protocol could be deployed or developed:
- Distributed Architecture[2]
- Data Centric Approach of Data Transmission [2] [3].
- Auto Network Configuration [2].
- Energy Efficient [2] [4].
- Reliable Routing Architecture [2].

The solution talks of having a protocol having above mentioned features as key ingredient. The features will be inhibited as follows.

*A. Distributed Architecture*

The prime need of such a network is Distributed Architecture. If data is transferred from one central node, due to lack of power or node damage if the node goes down, the network may collapse. Therefore, so as to increase the reliability of the network, a distribution in control may highly increase the reliability and stability of the network. To this, we need to put a little or no effort as the distribution can be easily achieved by selection of nodes based upon energy and probability of Stability in network. These set of nodes may be further used to transfer the data to selective cluster they belong to. For the selection method we may use the MPR node selection pattern as in OLSR protocol of Ad-Hoc Networks and we can accordingly modify the algorithm of selection as per the need of the environment or usage. The distributed control will result in:

- Back-up control [6].
- Less Network Failure Probability [6].
- Better Data Collection [6].

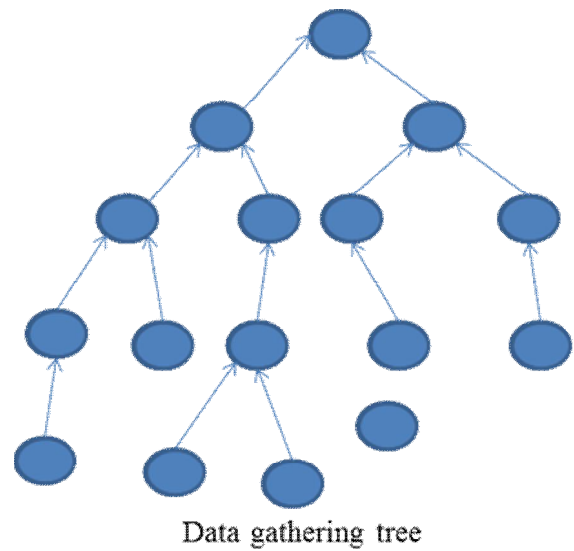

Figure 3: A data centric network flow

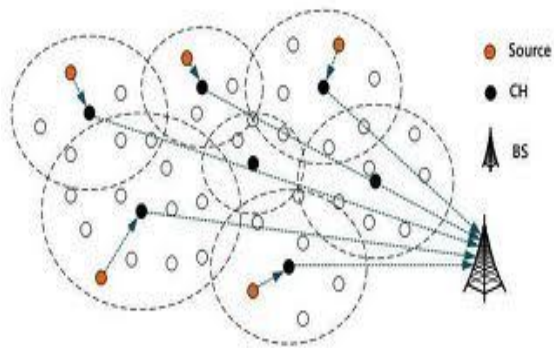

Figure 2: A Distributed architecture approach

*B. Data Centric*

When we talk of Wireless Sensor Networks, the first thought of most reliable transmission comes with flooding based routing. However, it does guarantee the maximum reliability but not 100 % guarantee of data delivery. Also, Flooding is very inefficient in terms of energy consumption. To this also, the solution does lie in the Ad- Hoc Networks through means of routing. The network generated in terms of distributed architecture, may easily be used here to determine routes to every other node in the network.

In various application of sensor network it is not efficient to provide global identifier to each and every node because of having sheer number of node deployed. Dearth of global identifier and the random deployment make it difficult to select a specific set of sensor nodes to be queried. So data is often transmitted from each sensor node within the deployment region with some redundancy. Since, this seems to be infeasible in reference to energy consumption, routing protocols that will be able to have set of sensor nodes and uses data aggregation [6] during the data relying must be considered. This consideration leads to data centric routing [3] which is different from traditional address based routing in terms of route created between addressable nodes managed in the network layer of communication stack.

On the other hand in the data centric routing sink send queries to specific region and waits for data from the sensors located in the particular region.

For the above purpose different protocols have been proposed on different theories.

*C. Auto Network Configuration*

To interchange the message with different node in the network there needs to be some mechanism that should be deployed. TCP/IP protocol makes it possible to communicate with different node of the network by associating a different IP address.

Since in mobile ad hoc network we don't have such centralized entity that could carry out this task. so the need of the protocol that can perform the task of network configuration in the dynamic or auto dynamic was essential. Since Mobile ad hoc network is dynamic topology in nature, auto configuration protocols faced with various problems in providing the uniqueness of IP address.

For the correct functioning of network the protocols strive to achieve the following features:-
- Assigning unique IP address
- Resolve the problems occur form loss of packet
- Allow multi-hop routing
- Minimization of additional packet traffic in the network
- Synchronization

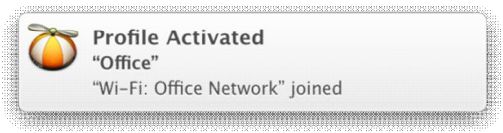

Figure 4: A message on auto Configuration

### D. Energy Efficient

Nodes in Wireless sensor network needs to be on continuous power supply which is practically not a possibility. To overcome this, the nodes need to implement an efficient method to increase lifetime of network. To this, many methods are present already. Few of the methods available are:
- Energy Aware routing [4].
- Implementing Multi-hop routing [4][5].

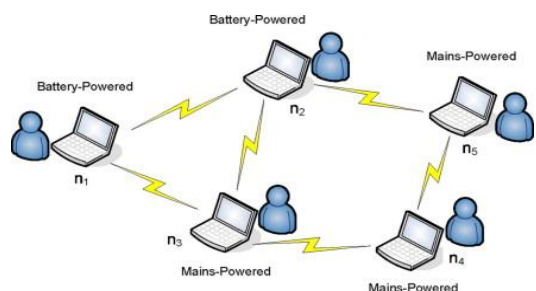

Figure 5: Energy Aware Routing

A network where energy could not be a constraint is not practically achievable in any Ad-hoc network. This could be accounted in terms of energy consumption by nodes in transmitting data. It is well understood, that a node spends maximum energy while transmission process rather than sensing or processing period.

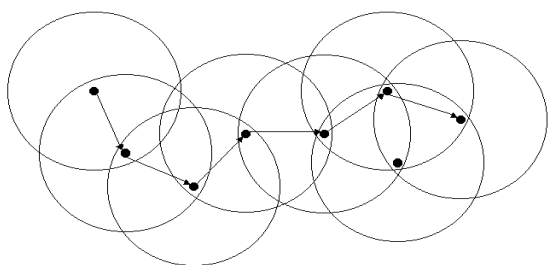

Figure 6: Multi-hop routing

In energy aware routing, while creating the infrastructure of the network, energy is the prominent constraint while designing the routes of transmission. While direct transmission to the destination node won't be a good idea based on the equation which clearly states that transmission power is proportional to the distance squared or even higher order if there are any obstructions present. Hence, multi hop routing is unavoidable. But, multi-hop routing also introduces significant overheads in terms of network topology management and data transmission management based upon the model of data transmission. This could put significant amount of power requirement over the nodes aggregating data while other nodes may be free.

### E. Reliable Routing Architecture

While working in an Ad-hoc mode, every node desires to have a reliable routing architecture in the backbone such that no failures, congestions, or other network problems persist. These requirements increase when we have a WSN as these networks are deployed where we require a prominent and efficient data transmission without 0% failure rate [7][8]. This is highly desirable and practically unachievable. To this, WSN protocols are present but when we have a Mobile Ad-Hoc Network (MANET), there occurs a special constraint of mobility, which in turn presents an issue of dynamically changing architecture of the network.

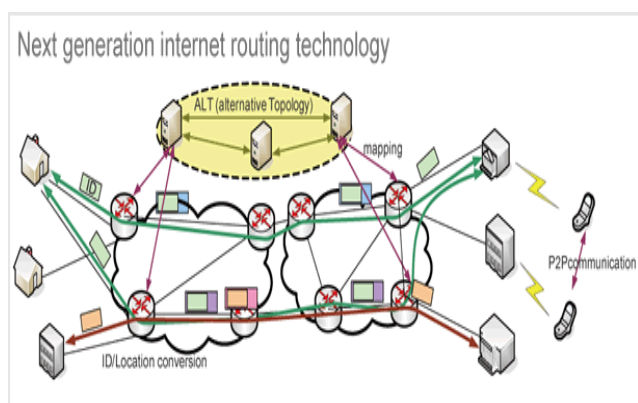

Figure 7: Reliable routing based on Alternate path availability

In order to be reliable in terms of routing by mobile WSN network, nodes must be capable of:
- Dynamic network updating based upon node movement [7].
- Adaptable to change configuration [7].
- Maintaining the core network management calls [7].

In addition to these, a quality highly desired is time bound operating scenario. These WSN nodes work in a scenario where time plays a very critical role. A loss in moment could lead to unimaginable outcomes which are not favourable.

### III. INTEROPERABILTY BASED SOLUTION

The issues discussed give us a clear view of requirements which are to be fulfilled by any protocol to be deployed in such scenarios. MANET protocols are quite sufficient in their working, but there are few point where there is a requirement to be fulfilled:

- Energy efficiency [7] [8].
- Data centric approach [7] [8].

These two are field where lots of research has been going on when it comes to routing protocols. Few of the solutions given by prominent researchers are:

- Maximum lifetime energy routing [12].
- Maximum lifetime data gathering [8] [12].
- Minimum cost forwarding [8] [10].
- SAR [8].
- SPEED [8] [9].

Out of the, aforesaid routing mechanisms the last two are specific to WSN protocol deployable. But, the top three methods are quite deployable in MANET routing scenario with few modifications. These methods are briefly explained in the following section.

#### A. Maximum lifetime energy routing

This is a very interesting mechanism presented by Chang and Tassiulas based upon flow of the network. The equation presented by them clearly presents an approach base upon defining link cost as a function of node remaining energy and the required energy to perform the transmission. It also directly relates to finding the traffic distribution or nodes present as a crowd in a portion of a network. The equation for finding the maximum lifetime energy is:

$$c_{ij} = \frac{1}{E_i - e_{ij}} \text{ and } c_{ij} = \frac{e_{ij}}{E_i}$$

#### B. Maximum lifetime data gathering

Kalpakis presented this mechanism for wireless sensor network protocols in terms of a polynomial time algorithm. This mechanism could be well deployed over a MANET routing protocol. It states that life 'T' of any system is defined as number of rounds or periodic data readings from the sensors until the first sensor dies. It has a schedule wherein a tree starts from the sink and spans to nodes in the network. The lifetime depends on duration for which every schedule remains valid. This directly point to the aim of maximizing the lifetime of the schedule. He also presented an algorithm called MLDA (Maximum lifetime Data Aggregation). The algorithm has an eye for data aggregation as prime constraint while setting up maximum lifetime routes.

#### C. Minimum cost forwarding

It is a two phased algorithm devised at finding minimum cost path in a large Wireless sensor network. It is not a flow based protocol and minimum cost path and resources on the node are updated at each flow end.

The first phase of protocol has back-off based algorithm wherein a message starts from sink and diffuses through network. Unlike flooding, every node adjusts its minimum cost only after retrieval of that single message in the flow.

In second phase, the source starts broadcasting data to neighbors. Nodes receiving this broadcasted message, adds its transmission cost to sink with the cost of the packet. If the cost is sufficient enough to reach the sink, packet is accepted and forwarded, otherwise dropped.

### IV. CONCLUSION

This paper presents how we can create a simple protocol for mobile wireless sensor network. There are many solutions available in form of well-known protocols like LEACH, SPIN etc. But, they are not tailor made for the mobile applications of WSN as referred at the beginning of the paper.

### V. FUTURE WORK

The future work could deal with designing an optimized protocol taking into considerations the issues for such a protocol and available solutions to them. It needs to be taken into account, which one suits best to the need of the scenario being taken care of. For eg, if we take battlefield scenario a protocol could be designed with basic WSN needs with all the requisites mentioned in Custom scenario section with Maximum lifetime data gathering approach customised for AODV routing or OLSR routing.